\documentclass{elsart}

\usepackage{epsfig}

\usepackage{amssymb}

\begin{document}

\begin{frontmatter}



\title{A First Mass Production of Gas Electron Multipliers}


\author[a]{P.S. Barbeau}
\author[a]{J.I. Collar}
\author[b]{J.D. Geissinger}
\author[c]{J. Miyamoto}
\author[c]{I. Shipsey}
\author[b]{R. Yang}

\address[a]{Enrico Fermi Institute and Center for Cosmological 
Physics, \\ University of Chicago, Chicago, IL 60637}
\address[b]{3M Co., Microinterconnect Systems Division, Austin, TX  78726}
\address[c]{Department of Physics, Purdue University, W. Lafayette, IN 47907}

\begin{abstract}
We report on the manufacture of a first batch of approximately 2,000 
Gas Electron 
Multipliers (GEMs) using 3M's fully automated roll to roll flexible 
circuit 
production line. This process allows low-cost, reproducible 
fabrication of a 
high volume of GEMs of dimensions up to 30$\times$30 cm$^{2}$. 
First tests indicate that the
resulting GEMs have optimal 
properties as radiation detectors. Production techniques and
preliminary measurements of GEM performance are described. This 
now demonstrated
industrial capability should help further establish the prominence of 
micropattern gas detectors in accelerator based and non-accelerator 
particle physics,  
imaging and 
photodetection.
\end{abstract}

\end{frontmatter}


A number of new radiation detector designs, collectively referred 
to as Micropattern
Gas Detectors
(MPGDs) \cite{reviews1,reviews2} have recently emerged in response to the 
extraordinary demands  
of next-generation High Energy Physics (HEP) experiments, namely the 
ability to respond to a 
high counting rate and integrated particle flux, 
superior radiation resistance and
fine spatial resolution. Common to
these designs is the presence of a large voltage drop 
(several hundred volts) across 
microstructures immersed in a suitable gas mixture.
Electrons originating from ionization of the gas in a conversion 
volume drift to the region of the microstructures where the 
intense electric field allows gas amplification to occur. Due to the 
confined amplification regions, slow positive ions are removed 
immediately from the amplification volume, increasing rate 
capability by several orders of magnitude compared to wire based gas detectors.

The attractive features of these detectors have lead to  
a growing number of applications in
many fields. For instance, MICROMEGAS chambers \cite{mms1} can be 
found nowadays
in medical digital X-ray imaging equipment \cite{eos}, where a
high sensitivity in low intensity radiation fields results in a diminished 
dose to the patient, while profiting
from an enhanced image contrast. Similarly, photocathode-coated MPGDs 
promise to surpass photomultiplier tubes in light detection efficiency, 
reduced cost and 
speed \cite{pmt1,pmt2}. Other emerging applications are industrial imaging 
\cite{indus} and X-ray 
astronomy \cite{costa}. Reviews of these can be found in 
\cite{reviews1,sharma}. 

Recently it has been proposed to extend MPGD use to 
the field of non-accelerator particle physics \cite{meyannis,ieee}, where 
uses would be numerous in view of their 
simplicity, the possibility to easily construct MPGDs out of radioclean materials 
and their very low energy threshold. It is in the context of the 
development of a new type of neutrino detector \cite{ieee} that we 
attempted to manufacture MPGDs in large numbers and with near-perfect
reproducibility, using an 
industrial approach. This effort may nevertheless have repercussions 
in satisfying the large demand for MPGDs in accelerator based physics. 
For this first attempt we 
chose a popular design, the 
Gas Electron Multiplier (GEM) \cite{GEMS} given its particular simplicity of 
design. 
A GEM consists of 
a $\sim 50 ~\mu m$-thick polyimide 
({\scriptsize  KAPTON}$^{\textrm{\scriptsize{TM}}}$) film copper clad on both sides, 
perforated with a regular matrix of small 
holes (diameter few tens of $\mu$m) produced by photolithography. 
When a voltage difference is applied between the two sides of the GEM, 
a large electric field is produced in the holes. Electrons that enter the holes 
undergo gas amplification. 
A remarkable
advantage of GEMs is the 
possibility of building multi-stage amplification layers \cite{ian}, 
where electrons are transferred from one GEM to the next, undergoing
successive avalanches and yielding very large charge gains. 
The resulting high-efficiency for
single electron detection looks particularly attractive to us, when considering the small 
energy depositions expected from low-energy neutrino recoils \cite{ieee}. 

\begin{figure}[tbp]
\epsfxsize = \hsize \epsfbox{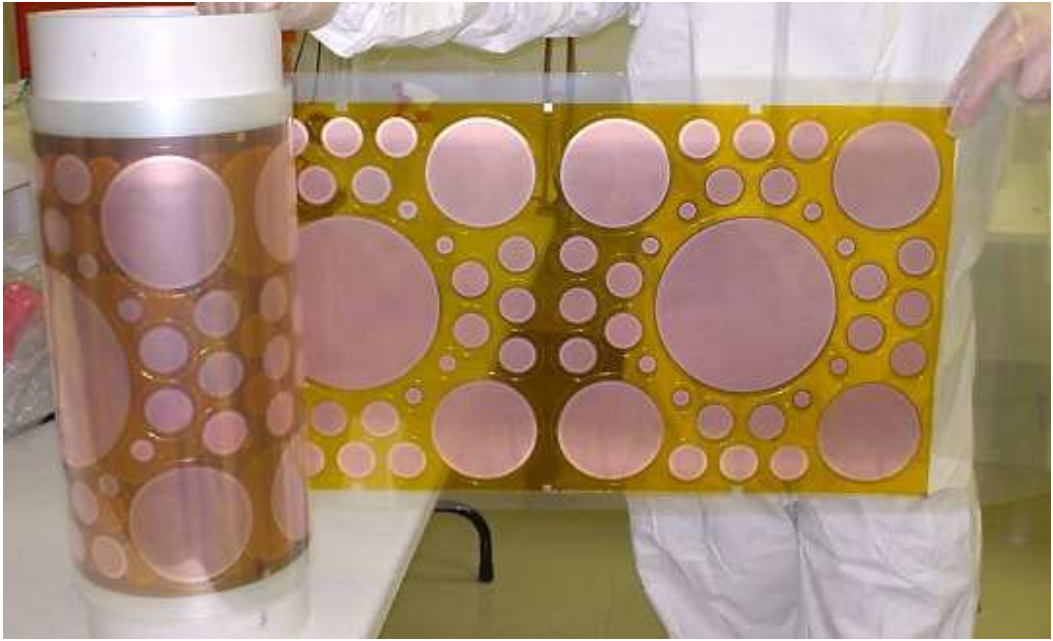}
\caption{Single continuous roll containing a production of $\sim1,000$ 
subtractive GEM 
elements in different sizes. Barely visible in the figure are  
perforations made 
around each GEM to facilitate detachment. The maximum 
GEM area permitted at present in 3M's production line is 30$\times$30 cm$^{2}$, already comparable 
to the largest MPGDs produced for high-energy physics experiments.} 
\end{figure}

\begin{figure}
\begin{center}
\includegraphics[width=.7\textwidth]{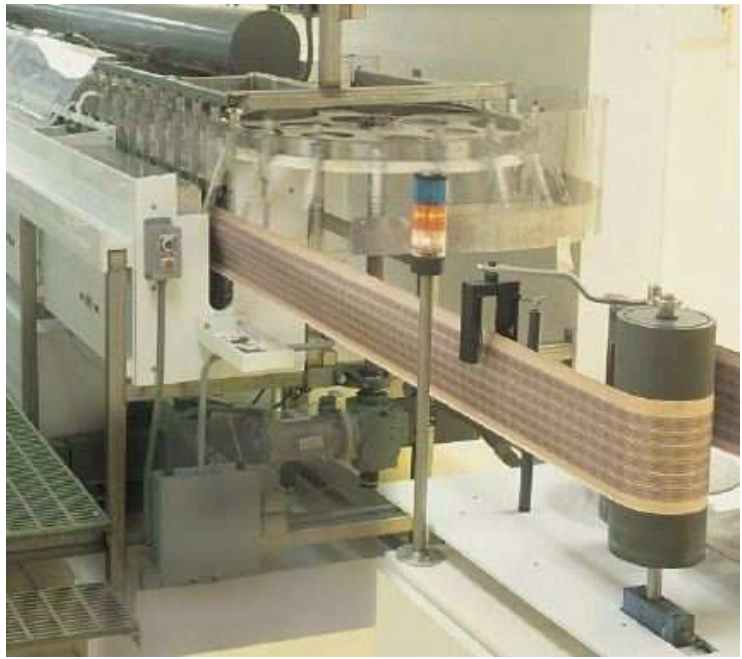}
\end{center}
\caption{3M's roll to roll flexible circuit manufacture in clean room 
conditions.}
\end{figure}

We report here on preliminary tests and observations made on a first 
batch of $\sim 2,000$ GEMS (\frenchspacing{Fig. 1}) produced 
using
3M's high volume, 
wide web, roll to roll, adhesiveless flexible circuit ({\scriptsize  
FLEX}) circuit making process 
(\frenchspacing{Fig. 2}). 
At the time of this writing every indication points at 
their having a satisfactory performance as radiation detectors. 
Our preliminary results are encouraging but testing is 
in an early stage. For example we have not yet studied the 
resistance to radiation (ageing) of the GEM.  We however feel 
that the widespread demand for GEMs by numerous research groups
justifies the early release of our findings.

Flexible circuits are utilized in a variety of applications 
such as inkjet printer cartridges, hard disk drives, liquid crystal display 
modules, and IC packages among others \cite{var}.  These applications have a 
variety of needs that are met with various {\scriptsize  FLEX}  
circuit constructions.  
These can be grouped into two categories:  3-layer and 
ÒadhesivelessÓ {\scriptsize  FLEX}  circuits \cite{ad}.  A 3-layer {\scriptsize  FLEX} 
is appropriately named 
since it is constructed from a copper foil, 
a polymeric film, and an adhesive to bond foil to film.  
In many applications such as Hard Disk Drives and GEMs, 
the presence of the adhesive would create outgassing and ionic problems. 
An adhesiveless {\scriptsize  FLEX}  circuit (also referred to as a 2-layer 
{\scriptsize  FLEX}) requires alternative 
means for securely bonding the copper to the polymer.  The two primary methods 
for fabricating an adhesiveless {\scriptsize  FLEX}  circuit are a) direct 
metallization of the polymeric film and b) casting of liquid polyimide onto the Cu foil.
After the substrate has been created, the copper and polyimide materials 
must be patterned to form the desired geometry for the application.  
The copper pattern can be formed by using either an additive or 
subtractive circuitization process. The  
process flows for each are  illustrated in \frenchspacing{Fig. 3}.

The additive process consists of applying a photo resist 
imaged with the desired copper pattern to a $50 ~\mu m$-thick 
polyimide film, directly metallized on 
both sides. The copper is then electroplated to the appropriate thickness onto the 
exposed flash layer.  This plating technique 
can allow for a wide range of copper thicknesses ranging from 4 to 36 $\mu$m.  
As shown in \frenchspacing{Fig. 4}, 
this additive circuitization process can achieve 
very fine copper features \cite{john},  down to $20 ~\mu m$ trace and $20 ~\mu m$ space on 
1-metal layer {\scriptsize  FLEX}  circuits ($30 ~\mu m$ traces and $30 ~\mu m$ spaces on 
2-metal layer {\scriptsize  FLEX}  circuits).  As can be seen in the trace cross section 
in the figure, the sidewalls on the additive copper are nearly vertical.

GEM foils were also manufactured using the subtractive process 
flow  outlined in \frenchspacing{Fig. 3}.  The subtractive structure is believed 
to be similar to the construction described by Bouclier {\it et al.} 
\cite{bou}. The side walls of the copper openings have a 
somewhat shallower slope than in the additive process.  
GEMs produced with these
two different circuitization processes 
showed some performance differences described below.

\begin{figure}[tbp]
\epsfxsize = \hsize \epsfbox{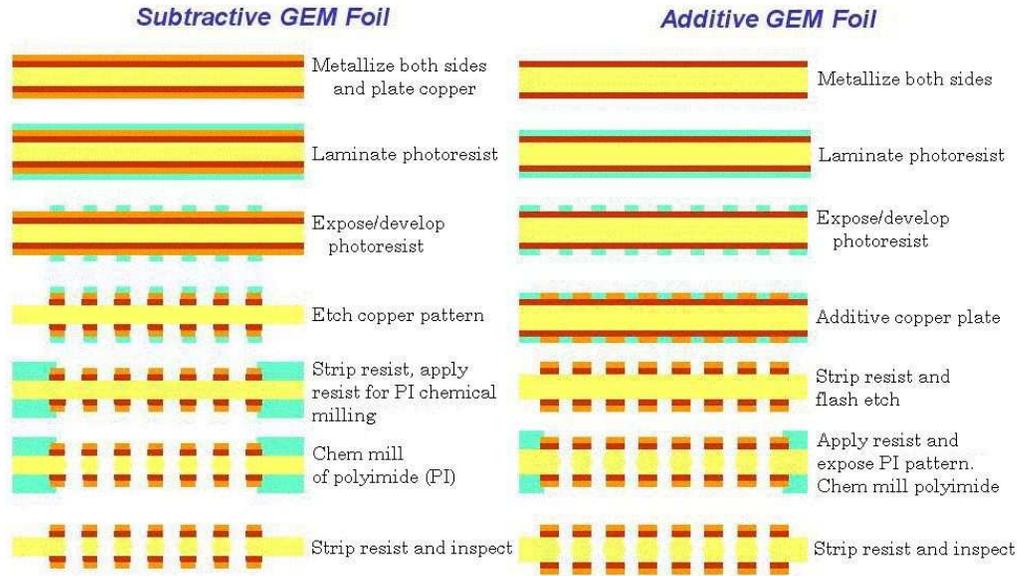}
\caption{Subtractive and additive process flows used in the manufacture 
of 3M's GEM foils.} 
\end{figure}

\begin{figure}
\begin{center}
\includegraphics[width=.7\textwidth]{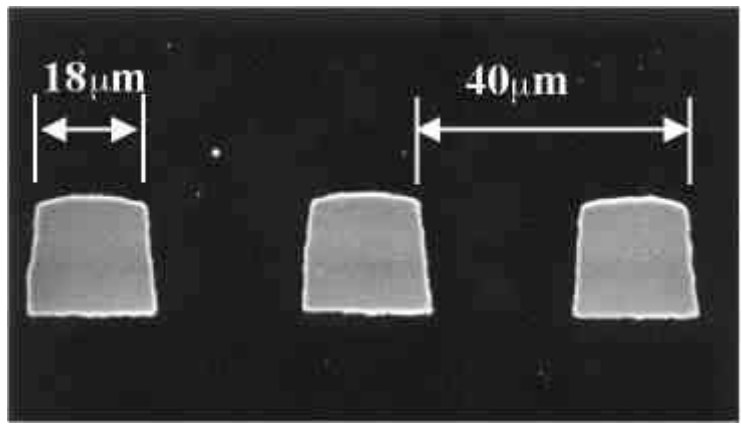}
\end{center}
\caption{$40 ~\mu$m pitch circuitization on 1-metal layer {\scriptsize  FLEX} 
(see text).}
\end{figure}

\begin{figure}[tbp]
\epsfxsize = \hsize \epsfbox{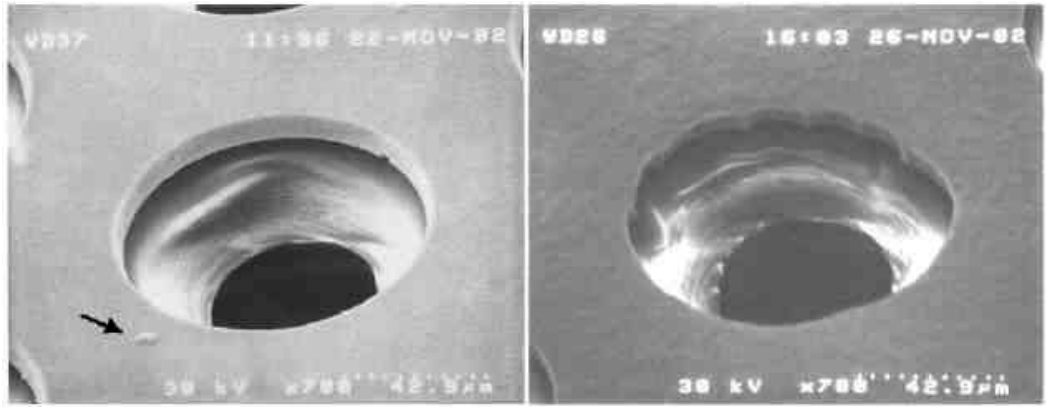}
\caption{Scanning electron microscope images of 3M subtractive (left) and 
additive GEMs (right). A small Cu microcrystal (height $<1 \mu$m) 
is indicated by an arrow on the subtractive surface. No sparking or 
other spurious effects have been observed from these. Additive GEMs 
display smoother surfaces but copper opening irregularities can be 
identified on large areas of the panels. 
An extreme 
case is depicted here (see text).} 
\end{figure}

\begin{figure}
\begin{center}
\includegraphics[width=.7\textwidth]{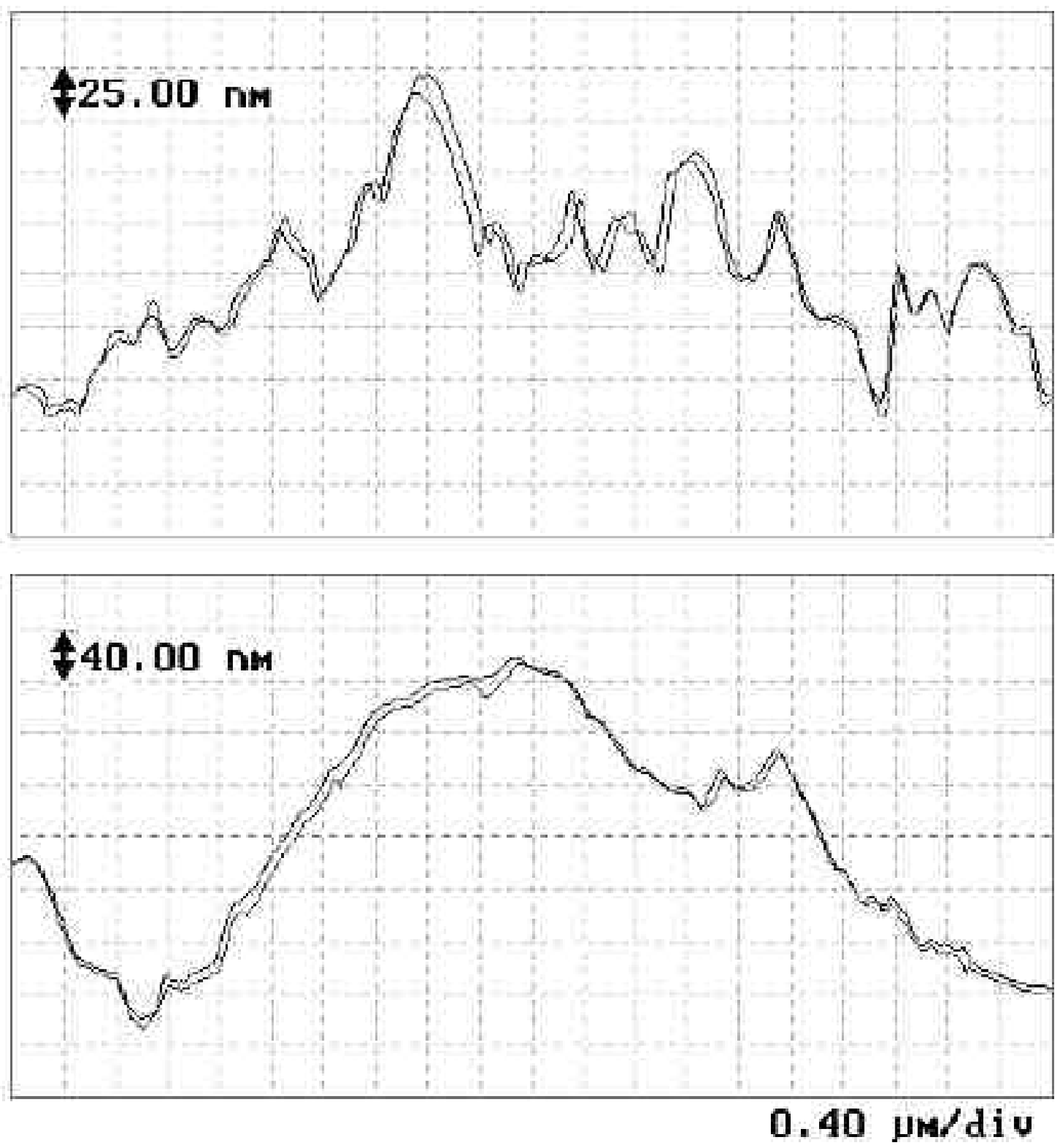}
\end{center}
\caption{Typical surface roughness in 3M's subtractive (top) and 
additive GEMs (bottom): the figures show tip traces and retraces
from contact-mode atomic force microscopy, scanning along a 
straight line. The smoother additive relief may be preferable in 
applications where field effect electron emission must be kept down to an absolute 
minimum. Note the difference in 
vertical scale.}
\end{figure}

One lot from each fabrication method has been produced so far, each 
containing $\sim$30 identical panels of 33 GEM elements (\frenchspacing{Fig. 
1}). Production of a much larger number of panels per lot, up to 
a few hundred, is possible.
In both cases the chosen design was the 
so-called ``standard GEM'' \cite{ben}, i.e., $80 ~\mu$m holes in an hexagonal 
pattern with $140 
~\mu$m pitch and a biconical transversal hole cross section. In other 
words, the innermost part of the 
holes exhibits a reduced opening of $\sim55 ~\mu$m, a 
characteristic also found in 
most GEMs built elsewhere. The use of Dupont E-film 
{\scriptsize  KAPTON}$^{\textrm{\scriptsize{TM}}}$ as the substrate 
does not allow to reduce this opening any further. In 
present lots the copper thickness was fixed at $12 ~\mu$m to insure 
the success of these first trials. In successive attempts this will be 
further reduced, a feature of interest for tracking devices where 
multiple scattering in the detector must be minimized. 

The surface quality of both lots has been studied via SEM
(\frenchspacing{Fig. 5}) and AFM (\frenchspacing{Fig. 6}). Slight
copper opening irregularities
are observable mostly on one side of these first additive  
GEMs. This may lead to gain inhomogeneities across the 
GEM surface: therefore we have concentrated at first on the 
characterization of the subtractive 
lot. The additive fabrication process has 
proven to be challenging: small polyimide ribs stemming from the interior of the 
holes were initially observed to 
envelope the edges of copper openings. Additional treatment of the 
lot removed these but resulted in a slightly diminished copper to 
polyimide attachment\footnote{It must be noted that the method used to 
test copper to polyimide attachment is probably too stringent, 
consisting of firmly attaching adhesive tape to the GEM surface and 
swiftly peeling it off. Only some additive 3M GEMs are seen {\it not} to 
pass the test. With any luck a GEM should not have to withstand such abuse 
during normal operation.}
and the previously mentioned irregularities. 
The smoother copper surface quality obtained 
with this method (\frenchspacing{Fig. 6}) 
is nevertheless a redeeming quality that justifies further exploration: it may be 
of importance in applications where total inhibition of field effect 
electron emission is sought, as is the case in \cite{ieee} and  
other efforts concerned with single-electron detection \cite{pmt2}. 
The subtractive surface quality 
exhibits apparently innocuous copper microcrystallite growths
(\frenchspacing{Fig. 5}): while we have not observed any sparking nor 
unexpected behavior from their presence, an attempt will be made to 
remove
them in new lots. In order to inhibit their growth it should suffice 
to reduce the concentration of 
dissolved copper in etchant baths.

\begin{figure}
\begin{center}
\includegraphics[width=.7\textwidth]{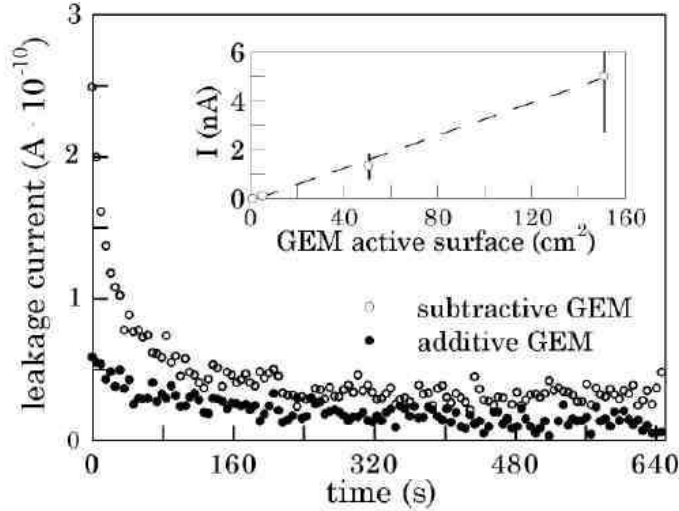}
\end{center}
\caption{Leakage current in air ($\sim 40\%$ humidity) across 5 
cm$^{2}$
additive and subtractive 3M GEMs. A 600V bias was
applied using tin clamps on their outermost 
$\sim 0.5$ cm (this annular region is devoid of holes to facilitate soldering). 
The measurements were performed with 
a Keithley 6485 picoammeter. The GEMs were enclosed in a special
double shielding to attenuate RFI/EMI interference \protect\cite{keit}.
{\it Insert:} Dependence of the  
leakage current (asymptotic value after several hours)
on subtractive GEM active 
surface area. A total of approximately twenty randomly-selected 
GEMs have been 
characterized, all displaying similar low values. The figure shows
averages and their 
dispersion.}
\end{figure}

\begin{figure}
\begin{center}
\includegraphics[width=.7\textwidth]{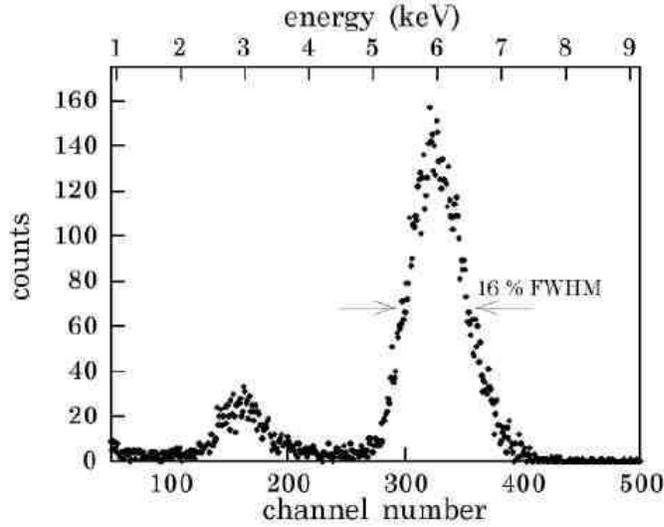}
\end{center}
\caption{Characteristic spectrum from an uncollimated $^{55}$Fe 
source and a single subtractive 3M GEM in Ar + 5\% 
CH$_{4}$ (active area 5 cm$^{2}$, V$_{drift}$ = 500V, V$_{GEM}$ = 
480V). The signals were read off the lower GEM electrode with 
a grounded PCB immediately beneath it to aid charge collection,
passed on to an
ORTEC 142AH low-noise preamplifier and recorded 
using a XIA POLARIS digital 
spectrometer.
Good energy resolution in the presence of an uncollimated source can be
an indicator of adequate gain 
uniformity across the surface. $^{55}$Fe resolutions 
down to $\sim 14$\% have been 
obtained from this lot in much less than optimal conditions (stagnant
gas, uncollimated source, $\sim 10$ cm drift length in an inhomogeneous  
drift field).}
\end{figure}

\begin{figure}
\begin{center}
\includegraphics[width=.9\textwidth]{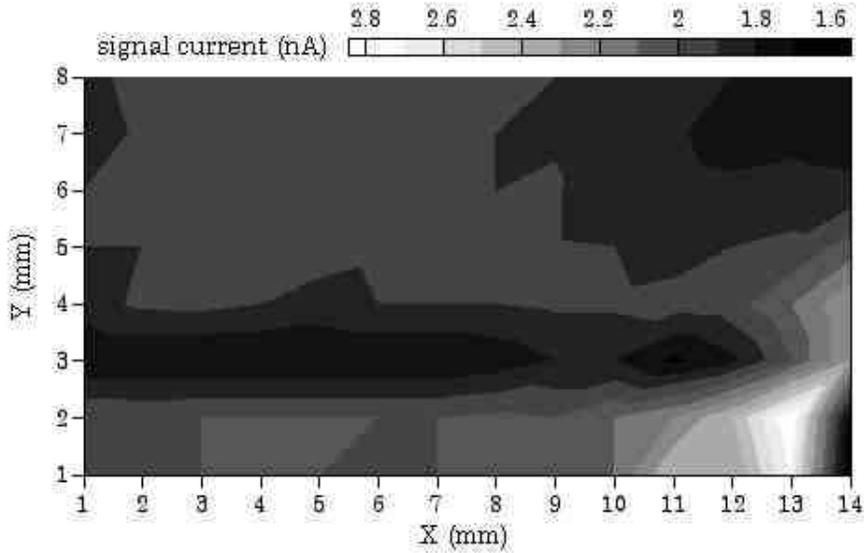}
\end{center}
\caption{Gain uniformity in Ar:DME (9:1) for a subtractive 3M GEM 
irradiated with a strong 5.4 keV
 X-ray source focused on a 1 mm$^{2}$ spot.
The current generated was measured with a picoammeter directly from 
the bottom GEM electrode (V$_{drift}$ = 600V, V$_{GEM}$ = 400V). 
The measured dispersion (standard deviation of 112 measurements) 
is 9\%.}
\end{figure}

\begin{figure}
\begin{center}
\includegraphics[width=.9\textwidth]{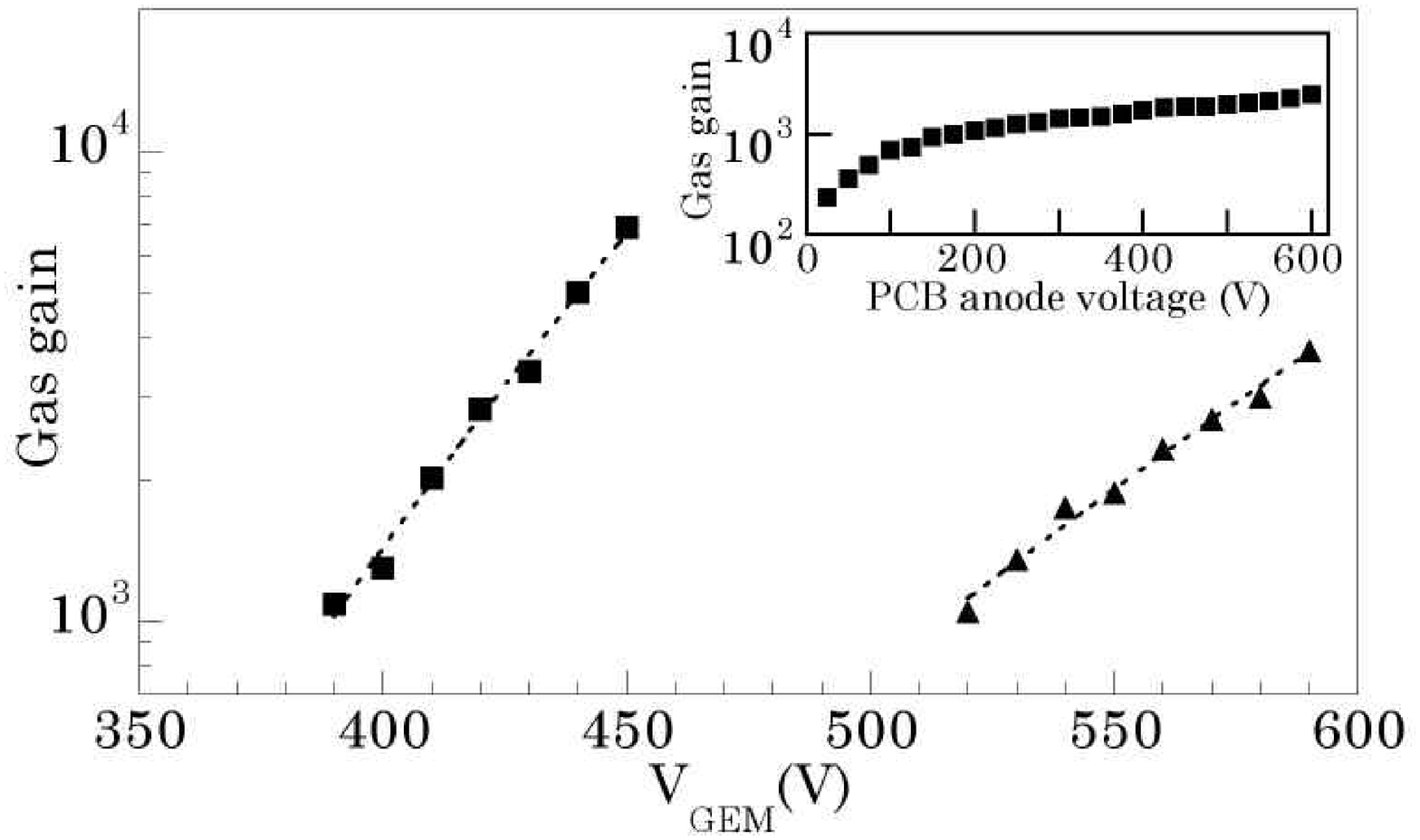}
\end{center}
\caption{Gain from an 8cm-diameter subtractive 3M GEM in Ar:DME (9:1, 
squares) and Ar:CO$_{2}$ (7:3, triangles) as a function of voltage 
across the element, in the presence of an uncollimated $^{55}$Fe 
source. Charge amplification was obtained with an ORTEC 142PC 
preamplifier collecting from the GEM lower electrode. The drift 
voltage was -500V. {\it Inset:} Gain in Ar:DME (9:1) 
using the more conventional approach of 
collecting from a single-channel PCB readout
placed 1 mm below the GEM (V$_{GEM ~UPPER}$ = -400V, 
V$_{GEM ~LOWER}$ = 0V), as a function of V$_{PCB}$.}
\end{figure}

\frenchspacing{Figs. 7-10} incorporate the extent of our preliminary 
characterization of subtractive 3M GEMs. 
\frenchspacing{Fig. 7} shows 
typical leakage currents measured in a number of randomly selected 
GEMs. They consistently display values comparable to previously 
produced GEMs. 
However, we have not yet observed any need to ``cure'' or ``burn'' 3M 
GEMs against shorts able to suddenly raise this current into the 
$\mu$A range, as is sometimes necessary with other GEMs. 
The good behavior of these leakage currents over periods 
of several hours probably comes
from the homogeneous surface treatment that the fully-automated 
roll to roll process guarantees, together with the use of high-purity 
polyimide, free of any 
fillers. Each part of each GEM foil receives an identical 
treatment in every fabrication step, something hard to achieve in 
manual production runs, especially over large surface areas.
For these same reasons we expect a good gain uniformity 
over large GEM surfaces.  An
optimal energy resolution in the presence of an uncollimated $^{55}$Fe 
source, as evidenced in \frenchspacing{Fig. 8}, points in this direction. 
As a matter of fact, first tests of gain uniformity 
(\frenchspacing{Fig. 9}) yield values already comparable to other 
MPGDs \cite{unif}.
Finally, \frenchspacing{Fig. 10} displays the gas gain measured using 
the GEMs as an isolated detector, i.e., without a charge collection 
backpanel anode. We observe no deviation from the expected behavior, nor 
any anomaly in the onset of discharges (at about V$_{GEM}$ = 450V in 
Ar + 10\% DME and 600V in Ar:CO$_{2}$). 

While the R\&D on these GEMs has barely started, all 
observations are presently very encouraging. First 
trials with a liquid crystal polymer (LCP) substrate show 
near-cylindrical hole walls, which can be of interest in applications 
where excessive dielectric charge-up via ion deposition is a concern 
(this can lead to a diminished gain uniformity across the surface).
Other advantages of 
LCP compared to {\scriptsize  KAPTON}$^{\textrm{\scriptsize{TM}}}$ \cite{lcp} are a much smaller maximum water absorption 
(0.02\% vs. 2\%, which may result in lower outgas, of 
relevance in HEP applications where extreme gas purity is required),
better
dielectric properties and a higher chemical and heat resistance.
The last may result in
GEMs more compatible with soldering and operation in commonly used 
etching detector gases such as CF$_{4}$, and possibly more resistant 
to sparking.
We expect to be able to report on LCP-GEMs soon. 

Hopefully the 
methods presented here will enable the production 
of large area MPGD's. These will be required in large next-generation 
time-projection chambers, the leading candidate for the tracking 
system at the next linear collider \cite{tpc} and also a possible contender 
in future underground physics experiments \cite{meyannis}. Proposals for hadron-blind 
GEM-based detectors in heavy-ion physics programs may similarly benefit 
\cite{aid}.

Tested GEM samples can be obtained from 
collar@uchicago.edu. JIC and PB would like
to thank Q. Guo for his assistance in 
performing SEM and AFM measurements and T. Witten for helpful 
discussions. JM and IPJS thank 
Kirk Arndt and Tom Smith of the 
Department for Physics at Purdue University for technical support.



\end{document}